\documentclass[prd,twocolumn,tightenlines,nofootinbib,amsfonts,amssymb,amsmath]{revtex4}
\usepackage{amsmath}\usepackage{amsfonts}\usepackage{graphicx}\usepackage{hyperref}\usepackage{mathbbol}

\def\be{\begin{equation}}   \def\ee{\end{equation}}   \def\bea{\begin{eqnarray}}    \def\eea{\end{eqnarray}}  \def\no{\nonumber}
    \def\d{{\rm d}}                
\def\f{\frac}

\def\l{\left}
\def\r{\right}

\def\p{\partial}

\begin{document}
\title{Thermodynamics Formulation of Economics}
\date{\today}

\author{Burin Gumjudpai}\email{burin.gum@mahidol.ac.th}
 \affiliation{Centre for Theoretical Physics \& Natural Philosophy ``Nakhonsawan Studiorum for Advanced Studies", Mahidol University, Nakhonsawan Campus,  Phayuha Khiri, Nakhonsawan 60130, Thailand}
 \affiliation{The Institute for Fundamental Study ``The Tah Poe Academia Institute", Naresuan University, Phitsanulok 65000, Thailand}
\affiliation{Thailand Center of Excellence in Physics, Ministry of Education, Bangkok 10400, Thailand}

\begin{abstract}
We consider demand-side economy. Using Caratheodory's approach, we define  empirical existence of equation of state (EoS) and coordinates.  We found new insights of thermodynamics EoS,  “the effect structure”.  Rules are proposed as criteria in promoting and classifying an empirical law to EoS status.  Four laws of  thermodynamics are given for economics.  We proposed a method to model the EoS with econometrics.  Consumer surplus in economics can not be considered as utility.  Concepts such as total wealth, generalized utility and generalized surplus are introduced. EoS provides solid foundation in statistical mechanics modelling of  economics and finance.
\end{abstract}
\maketitle

{\it This article is an extended abstract awarded the Richard Newbold Adams Medal for integration of natural and social sciences\footnote{https://iaisae.org/index.php/honors/}  at the International Conference on Thermodynamics 2.0, June 22-24, 2020 in  Massachusetts, USA.}
\section{INTRODUCTION}
There exist equilibrium states in economics and thermodynamics both of  which describe aggregated phenomena.  We criticize in how thermodynamical states are in existing. Key concepts are of the thermodynamics coordinate space and the equation of state (EoS).  Thermodynamics space is $2(n+1)$ tuples of $(X_i, Y_i, S, T)$  where $X_i, Y_i$ are extensive and intensive coordinates and $i = 1 \ldots n$.  At equilibrium $S$ is fixed,  the manifold becomes
$(2n+1)$ tuples, $\mathcal{M} = (X_i, Y_i, T) $. Existing of the manifold space is supported by empirical axioms
, the existence of thermal states and entropic states  (M\"{u}nster 1970  \cite{munster},  Land\'{e}  1926  \cite{cara}). These states relate two mechanical pair variables giving rise to the EoS,  $g(X_i, Y_i, T) = 0$.
For $n=1$ in hydrostatics system, this is $g(V,P,T)=0$ space. The EoS constrained the system to evolve on a 2-Dim surface. On the surface, thermodynamics potential as analogy to a two degree of freedom "field”.  For example, the internal energy, $U = U(X,Y)=U(Y,T)=U(X,T)$. With Legendre transformation, it can be transformed to other potentials. Energy transfer in thermodynamics comes in form of work, $\delta W = Y \d X$ and heat, $\delta \mathcal{Q} = T \d S$.  In economics, there are stock quantity or flow quantity. We should classify which stock variables  should be potential or should be coordinates. The flow is analogous to energy transfer.

There is no concept of EoS and potential in economics  (Debrue 1972 \cite{debrue}, Mas-Colell 1985 \cite{Mas}).  However similarities in common are: (1) equilibrium determined by a set of pair of dual state variables, (2) constraints of conservation, (3) preference of a quantity not to decrease in approaching equilibrium.  Following these similarities, Smith and Foley 2008 \cite{Smit} proposed $g(Q^{\d}, Pr, \rm{MRS}) = 0$ as EoS of consumers where $Q^{\d}, Pr, \rm{MRS}$ are demand quantity, price and marginal rate of substitution (among more than one type of commodity).
Previous attempts has long history back to Fisher's thesis (Fisher 1892 \cite{Fish}) which considered gradient  of utility with respect to  quantity of goods as analogy to
gradient of internal energy. Later, mechanical gradient of potential energy analogy to gradient of utility  with respect to quantity of goods was considered by Walras (Walras (1909) \cite{walras}). This had been
abandoned for many years until Samuelson’s critique  (Samuelson 1960 \cite{Samu}). There are more investigations by Lisman (Lisman 1949 \cite{Lisman}) and Saslow (Saslow 1999 \cite{Sasl}) in connecting thermodynamics variables to economics however the EoS was not considered therein.   A few aspects in comparison  are noticed as follow.

{\bf{In physics}}: (1) manifold as home of EoS and manifold coordinates are $(X, Y, T)$,  (2) prediction power,
(3) aiming for unification, (4) short relaxation time and controlled experiment, (5) potential (or field),
(6) more fundamental units  (meter, second, kg, Coulomb, Kelvin, mole).

{\bf{In economics}}: (1) no EoS and manifold coordinates are quantities of each type of goods $(Q^{\d}_j)$,
(2) empirical formulae with prediction power,
(3) unification is as crucial but separated by schools of thoughts, (4) long relaxation time - hardly attain equilibrium state (frequently with minor disturbance or big shocks),
(5) no concept of potential (or field), (6) fewer fundamental units  (money unit, time unit,  quantity of goods, number of agents (consumer, seller, company, etc.))

Systems or agents  of both subjects  have  (1) name, (2)
properties (elasticity), (3)  duty according to its properties (governed by laws), (4) systematic uncertainty (systematic risk). However only in social system or agents that possess ability to make
choice or option which is based on many factors such as memory, preference etc.

 Here we propose new idea to thermodynamics, the {\it effect structure} of an EoS and apply this to economics, such that  econometric modeling is to be guided by EoS effect structure idea.
We hope to propose a concise and more complete picture of “thermodynamical paradigm” of economics (see also  Gumjudpai 2018 \cite{BG1}, Gumjudpai and Sethapramote  2019 \cite{BG2} and Gumjudpai and Sethapramote  2019 \cite{BG3}).

\section{TRULY ENDOGENOUS FUNCTION AND EFFECT STRUCTURE DIAGRAM}
In this section, we develop concept of the truly endogenous function and effect structure diagram as  machinery tools for inductive reasoning toward generalized rules or statement for an empirical equation to attain a status of EoS.
This can be done by noticing common nature of relevant laws and theories.   In simple EoS such as ideal gas law, there are directions of effect as concluded. (a) firstly,  pressure $P$ can affect temperature $T$ or volume $V$ directly whereas  initial change in $P$ comes from externality.
(b) secondly, $T$ can affect $P$ directly whereas initial change in $T$ is from externality. More considerations are that $T$ can not affect $V$ directly, but only via $P$ and that $V$ can affect neither $T$ nor $P$ at all.
Change in $V$ can not be done by other external variables but can only be done by $P$, hence $V$ is always passive.  Effects from externality is  dubbed exogenous.  Only $T$ and $P$ can take the exogenous effect.
 Functions related internal EoS variables $(P, V, T)$ is endogenous functions.  Only the functions represent  the effects in (a) and (b) are {\it truly endogenous functions}. When writing $V = V(P,T)$, it looks fine. However in this consideration, it is in fact $V = V \circ P(T) =  V(P(T))$. Hence $T$ is exogenous for $V$.  Therefore, for a hydrostatics system, the truly endogenous  functions (denoted with tilde sign) are:
 \be
V = \tilde{V}(P), \;\;\;\; \;\;\;\;   P = \tilde{P}(T), \;\;\;\; \;\;\;\;    T = \tilde{T}(P)\,.
 \ee
Arguments of functions are the causes and the values of function are the effects.
\subsection{Class I and Class II Diagrams}
\begin{figure}[t]  \begin{center}
\includegraphics[width=6.0cm,height=2.4cm,angle=0]{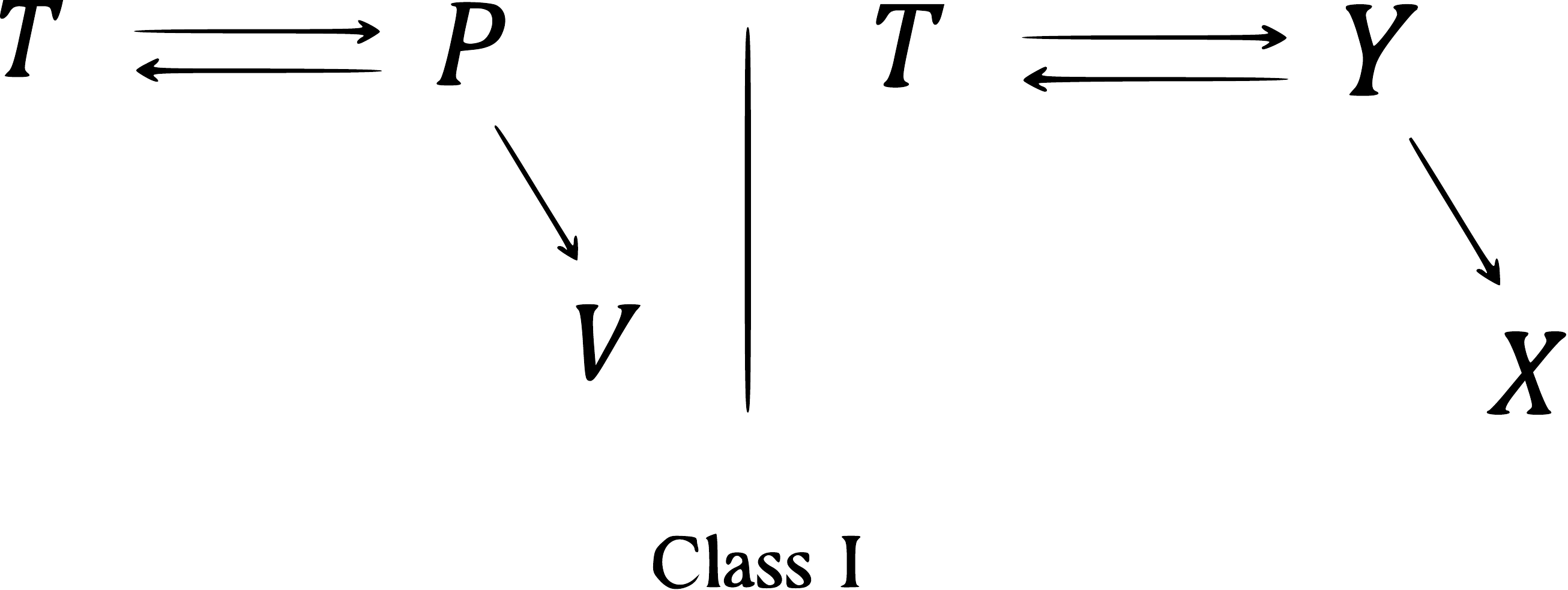}  \end{center}
\caption{EFFECT STRUCTURE DIAGRAM OF A CLASS I EOS: EXAMPLE  IS $(V,P,T)$ SYSTEM \label{figClass1}}
 \end{figure}

\begin{figure}[t]  \begin{center}
\includegraphics[width=6.0cm,height=2.4cm,angle=0]{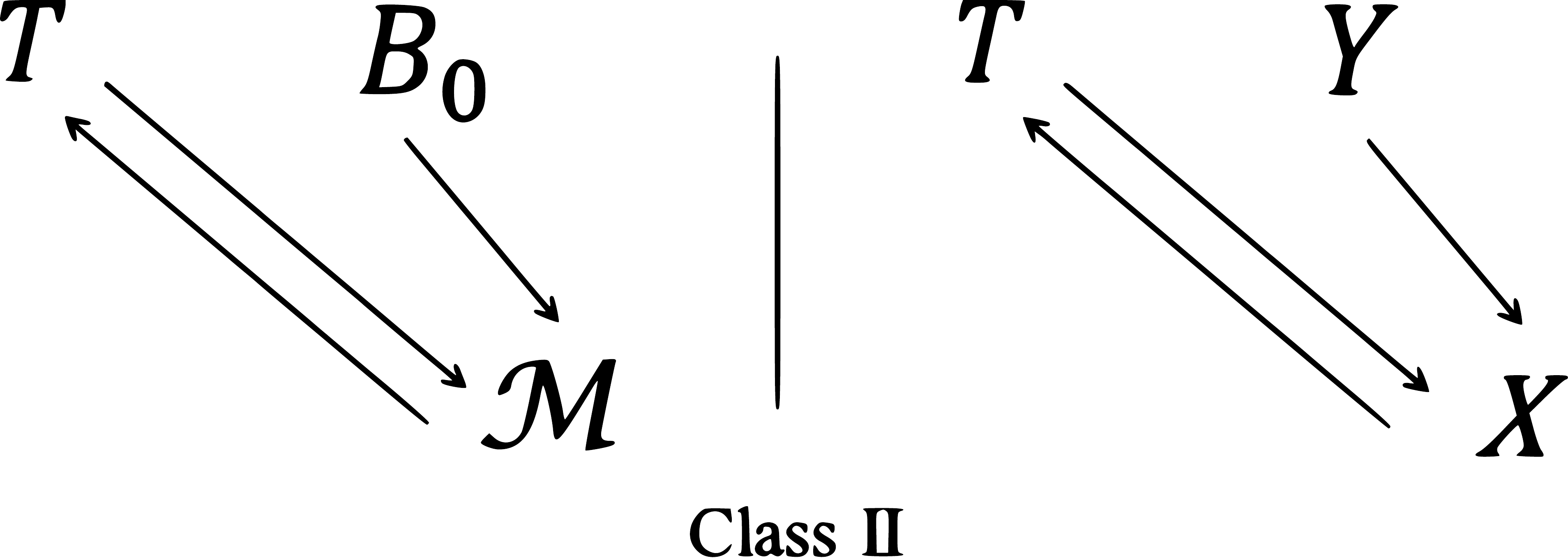}  \end{center}
\caption{EFFECT STRUCTURE DIAGRAM OF A CLASS II EOS: EXAMPLE IS  $(M,B,T)$  SYSTEM \label{figClass2}}
 \end{figure}

These  truly endogenous  functions can be represented as directed graphs in Fig. \ref{figClass1}. Arrows represent truly endogenous function as they  represent direction of cause and effect.  Initial changes in $T$ or in $Y$ are from exogenous effects, i.e. not from the variables within the diagram.
 We shall call the diagram with two opposite arrows linking $T$ and $Y$, the Class I diagram of EoS.
 Considering a paramagnetic substance\footnote{The case of paraelectric material is similar. Intensive quantity is electric field intensity and expensive quantity is electrical polarization.}, the EoS is $M =C B_0 / T$  where $M, C, B_0$ are magnetization (extensive), Curie constant and external magnetic field intensity (intensive). $T$ and $M$ can affect each others however $B_0$ can not be affected by neither $M$ nor $T$ as shown in  Fig. \ref{figClass2}.  The diagram is with two opposite arrows linking $T$ and $X$ hence it shall be realized as Class II diagram of EoS.

\subsection{A Proposal of Effect Structure Diagram Rules for the EoS Status}
After observation of many EoS in physical nature, we inductively form a set of rules as criteria for judging  status of an EoS, $g(X,Y,T) = 0$.  These are
\begin{enumerate}
  \item Number of the arrows is three.
  \item There is at least one arrow pointing $Y \rightarrow X$.
  \item Only $T$ and(or) $Y$ (apart from causing truly endogenous effects) can also taken exogenous influence or a shock. $X$ can not take exogenous effect unless via $Y$ or $T$  only\footnote{$X$ (e.g. volume) can be independent variable in the equation that related $Y$ and $T$, i.e. $Y=Y(T(X)) =  Y\circ T(X)$ or $T=T(Y(X)) = T\circ Y(X)$  so that changing $X$ might look like exogenous effect. However for a system in reality, given a value of $X$ at beginning, one can not change value of $X$ unless with truly endogenous effect from $Y$ and $T$.}.
\end{enumerate}

\begin{figure}[t]  \begin{center}
\includegraphics[width=6.0cm,height=2.4cm,angle=0]{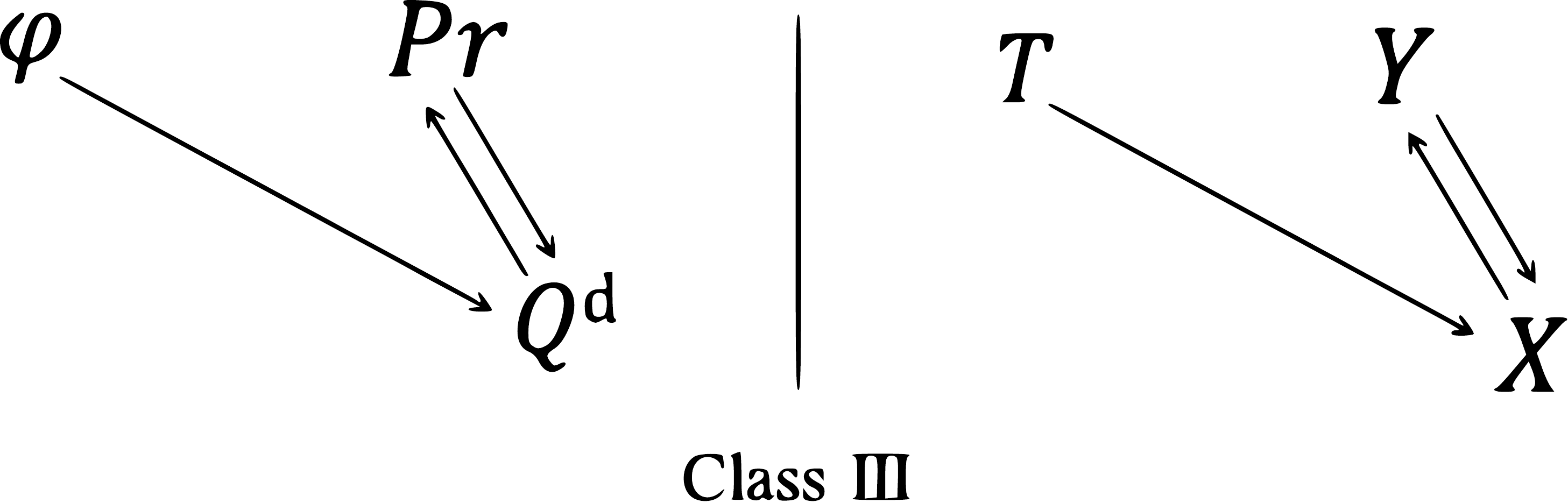}  \end{center}
\caption{EFFECT STRUCTURE DIAGRAM OF A CLASS III.1: DEMAND-SIDE ECONOMY $(Q^{\d}, Pr,\varphi)$ \label{figClass3}}
 \end{figure}

\section{DEMAND-SIDE  ECONOMY}
Assuming a one commodity market in perfect competitive condition with information symmetric and market clearance at equilibrium, we propose demand quantity $Q^{\d}$ as $X$ coordinate,
price $Pr$ as $Y$ coordinate and average personal wealth $\varphi$  as $T$.  In reality, $Q^{\d}$ takes passive role and is extensive with unit of number (amount) of goods. Typically, intensive coordinate $Y$ has a nature of force or influence per unit area or unit entity such as unit of charge. Here $Pr$ is the  cost of good per unit good, hence has money unit per unit of good.
For ideal system, $T$ is in fact  proportional to internal energy per mole or per particle, e.g. $T = 2U/(3 N k_{\rm B})$  in monatomic ideal gas. Hence $T$ is related to the variable defined by the first law, that is the internal energy, $U$.
The unit of energy is analogous to unit of money in economics. Therefore total wealth function $\mathcal{W}$ (in money unit) should play a role as $U$. As a consequence, average personal wealth $\varphi$ which plays a role of temperature should be defined as
\be
\varphi \;=\; \f{\mathcal{W}}{N}
\ee
where $N$ is number of consumers in the system.  The  total wealth function $\mathcal{W}$  comes in two parts. It is a combination,
\be
\mathcal{W}  \;= \; {\text{wealth\;\,in\;\,happiness}}\;\; +\;\; {\text{wealth\;\, in\;\,asset}}\,.
\ee
As this setting,  $ \mathcal{W}$ is a field potential,
\be \mathcal{W} = \mathcal{W}(Q^{\d}, Pr, \varphi) \ee.
To strictly define space $(Q^{\d}, Pr, \varphi)$, one needs to consider Carathéodory’s axioms which consider empirical existence of a set of infinite “thermal”  states,
\be
f(Pr_1, Q^{\rm d}_1)  = \varphi_1  \;\;\; \text{and} \;\;\; f(Pr_2, Q^{\rm d}_2)  = \varphi_2 \;\;\; \text{and\;\,so\;\,forth}\,.
\ee
and  existence of a set of infinite “entropic”  states,
\be
s(Pr_1, Q^{\rm d}_1)  = S^{\d}_1  \;\;\; \text{and} \;\;\;s(Pr_2, Q^{\rm d}_2)  = S^{\d}_2 \;\;\; \text{and\;\,so\;\,forth}\,.
\ee
This allows EoS $g(Q^{\d}, Pr, \varphi) = 0$ to exist. The EoS reduces one degree of freedom of $\mathcal{W}$. Hence  total wealth function is defined only on the EoS 2-Dim. surface.

Mechanical changes in total wealth of consumers is defined by work term (demand-side) of mechanical pair” $(-Pr, Q^{\d})$. The work term
\be \delta W^{\d} =  -Pr\, \d Q^{\d} \ee
is the expenditure for the generalized utility $\delta \mathcal{Q}^{\d}_{\rm util}$  which is a heat term,
\be
\delta \mathcal{Q}^{\d}_{\rm util}   = \varphi \, \d S^{\d}
\ee
for reversible processes.

 Here, unlike  microeconomics, maximizing of generalized utility $\delta \mathcal{Q}^{\d}_{\rm util}$  is not derived  by optimizing  with respect to quantity of different types of goods, $Q^{\d}_j$  (economics space coordinates) under a budget constraint. Instead generalized utility can be maximized with the second law to be stated later.  This is achieved  even for one commodity consumption from expenditure or it can be achieved even without  expenditure $\delta W^{\d}$, but only  with $\d S^{\d} \geq 0$.  Concept of generalized utility (heat term) is the combination,
 \bea
\delta \mathcal{Q}^{\d}_{\rm util} \;= &&   \delta({\text{pleasure\;or\;opportunity\;of\;ownership\;in\;assets}}) \no \\ &&+ \delta (\text{happiness\;\, of\;\.utilizing \;commodity})
 \eea
and the change in $S^{\d}$ is hence
\be
\Delta S^{\d} \geq \int \f{\delta \mathcal{Q}^{\d}_{\rm util}}{\varphi},   \ee    i.e. change in generalized utility per unit of average personal wealth.
\subsection{Class III Diagram}
Following effect structure diagram rules, and  with observations of fact in economy, the effect structure diagram (Fig. \ref{figClass3}) for demand-side economy is of a new class with two opposite-direction arrows linking $Y$ and $X$, the Class III.1. The rules also allow Class III.2-4 of which their effect are as follow: $(X \rightarrow T)$, $(T \rightarrow Y)$ and $(Y \rightarrow T)$.
\subsection{The Zeroth Law}
The zeroth law gives the existence of average personal wealth $\varphi$. Although hard to measure in society, however it helps defining personal wealth equilibrium when there is thermal (wealth) contact.
 Two consumers with different personal wealth can share via marriage, partnership, adoption, becoming family or as one household. Equal personal wealth should be approach but very slowly.
 Generalised utility is transferred in the contact (all in money unit).  Assets ownership (a sector of generalized utility) can be transferred with or without expenditure, i.e. no work term.
\subsection{The First Law}
The first law gives existence of total wealth function $\mathcal{W} = \mathcal{W}(Q^{\d}, Pr, \varphi)$ with conservation,
\be
\d \mathcal{W}  \;=\;  \delta \mathcal{Q}_{\rm util}^{\d}   + \delta W^{\d}
\ee
\subsection{The Second Law}
The second law provide existence of entropic function $S^{\d}$    interpreted as generalized utility per unit of personal wealth with,
\be
\Delta S^{\d}  \geq  \int \l({\delta \mathcal{Q}_{\rm util}^{\d}}/{\varphi}\r)\,.
\ee
In any processes of demand side, utility/(unit of personal wealth) does not decrease.  Meaning  of   $S^{\d}$    as utility/(unit of personal wealth)  can be understood from the example.
Rich people (high personal wealth) have less happiness in using a product.   It is cheap for them to buy the product.
\subsection{The Third Law}
At zero personal wealth, utility per unit of personal wealth  is zero. It takes infinite steps to reach zero personal wealth. Wealth includes happiness, there is no way to take it away completely.
If zero personal wealth exists, i.e. no happiness in any form, there is zero utility for the consumer.

\subsection{Processes in Demand-Side  Economy}
Adiabatic process (zero heat ),
\be \d \mathcal{W} = \delta W^{\d} = - Pr\, \d Q^{\d} \ee
 means that there is a spending without gaining any utility.
For an economics version of isovolumic process (zero work),
\be \d \mathcal{W} = \delta \mathcal{Q}^{\d}_{\rm util} =  \varphi \, \d S^{\d}\,.   \ee
Idealistically this is a gaining of utility without spending.
Considering isothermal process, there is group of consumers with constant average personal wealth.
Processes are governed by Smith’s price mechanism, i.e.  law of demand. Hence microeconomics is a specific case of our theory.
For the last case of isobaric process, i.e. constant price, demand quantity depends on only average personal wealth. If averagely consumers are richer from exogenous effect, demand should increase.

\subsection{Generalized Consumer Surplus, Total Generalized Utility and Total Expenditure}
\begin{figure}[t]  \begin{center}
\includegraphics[width=7.5cm,height=4.5cm,angle=0]{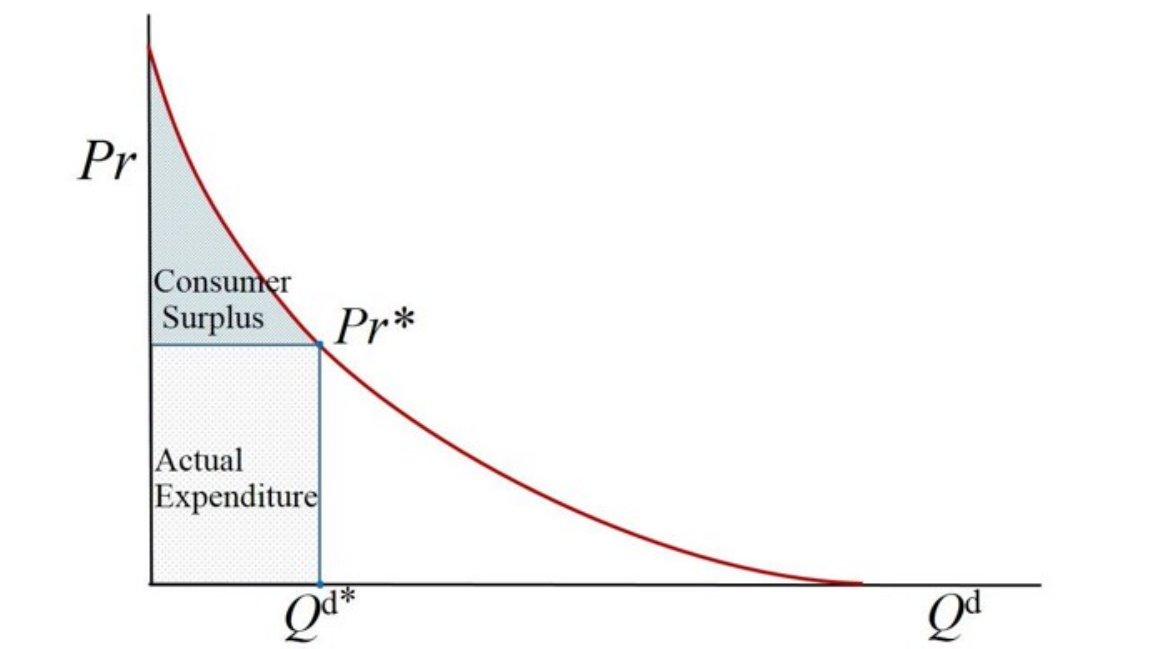}  \end{center}
\caption{CONSUMER SURPLUS IN ECONOMICS DEMAND CURVE   \label{fig4}}
 \end{figure}

\begin{figure}[t]  \begin{center}
\includegraphics[width=7.5cm,height=4.5cm,angle=0]{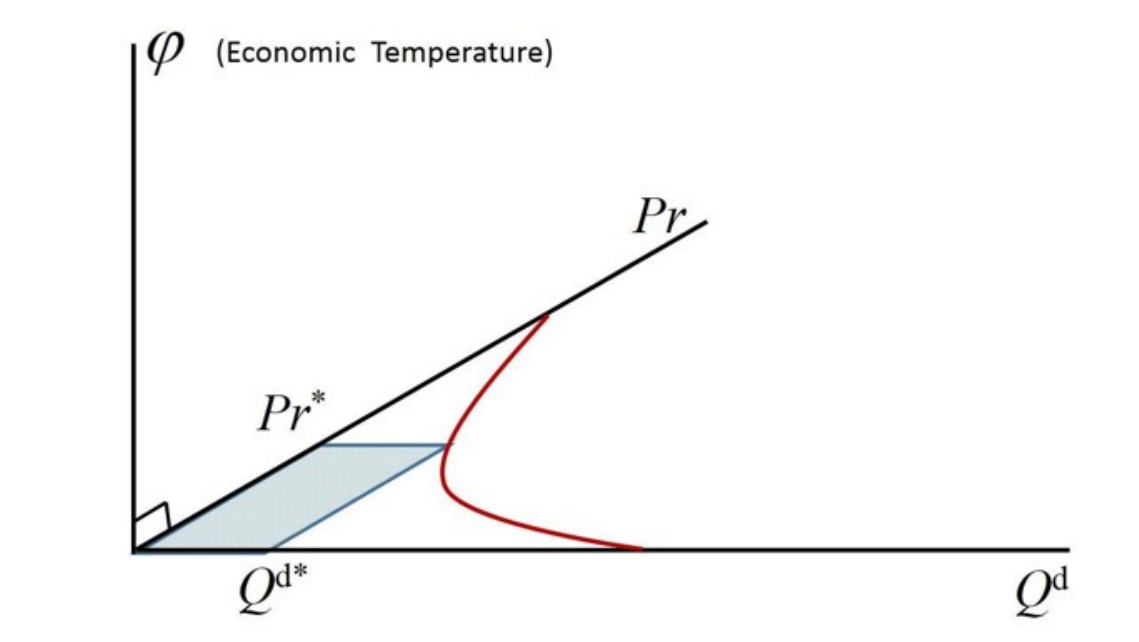}  \end{center}
\caption{THERMODYNAMICS-SUGGESTING SPACE FOR DEMAND-SIDE ECONOMY    \label{fig5}}
 \end{figure}

Consumer surplus in economics does not fit in our finding (see Fig. \ref{fig4}). The surplus should not be interpreted as utility and it should be picturized  beyond 2-Dim $( Pr,Q^{\d})$ plane
 into the $\varphi$ axis (see Fig. \ref{fig5}). Utility (heat) is not the size of area under curve in the mechanical pair plane, instead  utility should be the product of $\varphi$ and $S^{\d}$.
 At equilibrium (denoted with $*$), the total generalized utility is $\varphi^* \,S^{\d*}$. We suggest that the
generalized consumer surplus should be defined as
\be \Psi  \equiv   \varphi^*  S^{\d *}     - Pr^*\, Q^{\d *}, \ee
which is the difference between total generalized utility and total expenditure at equilibrium.


\section{MODELLING DEMAND-SIDE  EOS}
EoS in general form is  $g(Q^{\rm d}, Pr, \varphi) = 0$. With total differential method, we shall express,
\bea
\d Q^{\rm d}  & = &   \l(\f{\p Q^{\rm d} }{\p \varphi}   \r)_{Pr} \d \varphi   +   \l(\f{\p Q^{\rm d} }{\p Pr}  \,. \r)_{\varphi} \d Pr\,.  \\
\beta_{Pr}  & \equiv &   \f{1}{Q^{\rm d}_0}  \l(\f{\p Q^{\rm d}}{\p \varphi}  \r)_{Pr}  \; = \;  \f{E^{\rm d}_{\varphi}}{\varphi_0}\,, \\
\kappa_{\varphi}  & \equiv &   -\f{1}{Q^{\rm d}_0}  \l(\f{\p Q^{\rm d}}{\p Pr}  \r)_{\varphi} \; = \;  -\f{E^{\rm d}_{Pr}}{Pr_0},
\eea
where $E^{\rm d}_{\varphi}$ and  $E^{\rm d}_{Pr}$ are elasticities of demand to personal wealth and demand to price. In simplest case, elasticities may be assumed constant\footnote{Elasticities could depend on other factors as similar to the factors in EoS of simple solid.}.   It is straightforward  to write,
\be
Q^{\rm d}(Pr, \varphi) \;=\;  Q^{\rm d}_0 \l[ 1 + \beta_{Pr}(\varphi - \varphi_0)  -  \kappa_{\varphi}(Pr - Pr_0)  \r] \,,  \label{eqbgtbgtkk}
\ee
which is the EoS for the demand-side system.
Further we define $
Y  \equiv  {Q^{\rm d}  }/{Q^{\rm d}_0}\,, \;\; \;
X_1  \equiv  \varphi - \varphi_0\,,   \;\;\;\;
X_2  \equiv  Pr - Pr_0\,.
$
and our econometric model for the EoS is hence,
\be
Y \;=\;    1 + \beta_{Pr}X_1  -  \kappa_{\varphi} X_2  +   u \,,
\ee
where $u$ is an error term and this can be done with regression analysis in econometrics.

\section{CONCLUSION}
We analyse nature of EoS and found effect structure diagrams and its rules for the EoS. We use these criteria to variables in demand-side economy with hopes to express thermodynamics formulation of economics and explore the analogies and interpretation in various aspects.

\section*{Acknowledgments}
The author thanks International Association for the Integration of Science and Engineering (IAISAE) for the prize and for the invitation to present this work at the Thermodynamics 2.0 conference.

\end{document}